\documentclass[useAMS,usenatbib]{mn2e}
\usepackage{graphicx}
\usepackage{times}
\usepackage{epsfig}
\usepackage{rotating}
\usepackage{sidecap}
\usepackage{longtable}
\usepackage{lscape}

\usepackage[usenames]{color}

\def\src{XSS J12270--4859}

\def\gsrc{2FGL 1227.7--4853}
\newcommand{\is}{IBIS/ISGRI }
\newcommand{\je}{JEM--X }

\title[A propeller scenario for {\src}]{A propeller scenario for the
  gamma-ray emission of \\ low-mass X-ray binaries: The case of {\src}} \author[Papitto, Torres,
  Li]{A. Papitto$^{1}$\thanks{E-mail: papitto@ice.csic.es},
  D.~F.~Torres$^{1,2}$, Jian Li$^{1}$\\ $^{1}$Institute of Space
  Sciences (IEEC-CSIC), Campus UAB, Fac. de Ci\`encies, Torre C5,
  parell, 2a planta, 08193 Barcelona, Spain \\$^{2}$ Instituci\'o
  Catalana de Recerca i Estudis Avan\c{c}ats (ICREA), 08010 Barcelona,
  Spain }
\begin{document}

\pagerange{\pageref{firstpage}--\pageref{lastpage}} \pubyear{}

\maketitle

\label{firstpage}

\begin{abstract}

\src\ is the only low mass X-ray binary (LMXB) with a proposed
persistent gamma-ray counterpart in the {\it Fermi}-LAT domain,
{\gsrc}.  Here, we present the results of the analysis of recent
INTEGRAL observations, aimed at assessing the long-term variability of
the hard X-ray emission, and thus the stability of the accretion
state.  We confirm that the source behaves as a persistent hard X-ray
emitter between 2003 and 2012.  We propose that \src\ hosts a neutron
star in a propeller state, a state we investigate in detail,
developing a theoretical model to reproduce the associated X-ray and
gamma-ray properties. This model can be understood as being of a more
general nature, representing a viable alternative by which LMXBs can
appear as gamma-ray sources. In particular, this may apply to the case
of millisecond pulsars performing a transition from a state powered by
the rotation of their magnetic field, to a state powered by matter
in-fall, such as that recently observed from the transitional pulsar
PSR J1023+0038. While the surface magnetic field of a typical NS in a
LMXB is lower by more than four orders of magnitude than the much more
intense fields of neutron stars accompanying high-mass binaries, the
radius at which the matter in-flow is truncated in a NS-LMXB system is
much lower. The magnetic field at the magnetospheric interface is then
orders of magnitude larger at this interface, and as consequence, so
is the power to accelerate electrons.  We demonstrate that the cooling
of the accelerated electron population takes place mainly through
synchrotron interaction with the magnetic field permeating the
interface, and through inverse Compton losses due to the interaction
between the electrons and the synchrotron photons they emit. We found
that self-synchrotron Compton processes can explain the high energy
phenomenology of \src.

\end{abstract}

\begin{keywords}
acceleration of particles -- accretion, accretion discs -- magnetic fields -- X-rays: individual: -- gamma-rays: stars
\end{keywords}

\section{Introduction}

When captured by the gravitation of a neutron star, the interaction
between the matter outflow coming from a companion star (such as the
Be decretion disc of a Be star, or the Roche Lobe overflowing matter
from a low mass star) and the magnetic field of a neutron star can
lead to several states. A pulsar (ejector), a propeller, or an
accretion state can be realised depending on the balance between the
pressure exerted by the inflowing matter and by the rotating magnetic
field of the neutron star \citep[see, e.g.][for a
  review]{lipunov1992}. When the mass in-flow is able to bound the
magnetosphere to a closed configuration, whether accretion down to the
neutron star surface is possible (accretor state) or mass is
propellered away by the neutron star magnetosphere (propeller state)
is mainly determined by the ratio between the rotation rate of the
magnetosphere and of the incoming matter at the magnetospheric
boundary \citep{illarionov1975}.  At such interface, in some cases the
magnetosphere yields energy and angular momentum to the matter inflow,
and the plasma is expected to be very turbulent and magnetised.
\citet{bednarek2009b,bednarek2009} argued how in such conditions
electrons can be accelerated to high energies by a Fermi process,
yielding a detectable emission at GeV and TeV energies. In a similar
context, \cite{torres2012} explained the seemingly anti-correlated
orbital variability of the GeV and TeV emission of LS +61$^{\circ}$
303 in terms of the alternation between the propeller and the ejector
state of a magnetised neutron star, as the neutron star experiences
different mass in-flow rates along its orbit.

Here, we propose a propeller scenario to explain the properties of the
high energy emission of {\src}, so far the only low-mass X-ray binary
(LMXB) proposed to have a persistent gamma-ray counterpart, actually
emitting a comparable power in X-rays and at HE, {\gsrc}
\citep{demartino2010,hill2011,demartino2013}. Unlike gamma-ray
binaries, in this case both the wind and the radiative output of the
low mass companion star are unimportant in determining the HE emission
properties. Instead, we propose a model in which electrons are
accelerated at the interface between an accretion disc and a
propellering neutron star, and which losses are mainly driven by
synchrotron emission. Indeed, this electron population interacts with
the magnetic field permeating such layer and with the radiation thus
produced to yield the X-ray and GeV emission observed from the source.

The paper is organised as follows. In Sec.~\ref{sec:xss} we review the
properties of {\src} and of its proposed $\gamma$-ray counterpart,
\gsrc. In Sec.~\ref{sec:igr} we present the results of the analysis of
recent INTEGRAL observations, aimed at assessing the long-term
variability of the hard X-ray emission, and thus the stability
    of the accretion state. In Sec.~\ref{sec:propeller} we derive
expressions relating the observed luminosity to the relevant physical
parameters of the system, spin period, magnetic field, and mass inflow
rate, under the assumption that it hosts a propellering neutron star
with typical parameters of LMXBs. In Sec.~\ref{sec:sed} we reproduce
semi-quantitatively the high energy spectral energy distribution
produced by a relativistic population of electrons located at the
boundary between an accretion disc and a propellering magnetosphere,
under simple assumptions on the shape of the emitting region.  We
discuss these results in Sec.~\ref{sec:disc}, comparing our scenario
with other possible models proposed to explain the system, involving
either a rotation powered pulsar or an accreting compact object.

\section{\src}
\label{sec:xss}

{\src} is a faint hard X-ray source, first identified as a Cataclysmic
Variable on the basis of its optical spectrum
\citep{masetti2006}. However, the absence of a clear modulation of the
emission in the X-ray \citep{saitou2009,demartino2010} and in the
optical bands \citep{pretorius2009}, as well as the absence of Fe
K-$\alpha$ features in X-rays, forced to disregard such
hypothesis.


{\src} was observed on several occasions in the X-ray band: by RXTE in
November 2007 and during 2011 \citep[3--60
  keV;][]{butters2008,demartino2013}, by XMM-Newton in January 2009
and January 2011 \citep[0.5--10 keV;][]{demartino2010,demartino2013},
by Suzaku in August 2009 \citep[0.2--12 and 10--600
  keV;][]{saitou2009}, by Swift/XRT between March and September 2011
\citep[0.5--10 keV][]{demartino2013}, and by INTEGRAL since March 2003
\citep[20--100 keV; see][who reported the analysis up to October
  2007]{demartino2010}. Its average 0.2-100 keV luminosity was
evaluated by \citet{demartino2010} as
$L_{X}=(2.2\pm0.4)\times10^{34}\,d_2^2\,\rm{erg s^{-1}}$, where $d_2$
is the distance in units of 2 kpc. Its spectrum is described by a
featureless power law, $F_{E}\propto E^{-(\Gamma_X-1)}$, with an index
$\Gamma_X=1.70\pm0.02$, without any detected cut-off up to 100 keV
\citep{saitou2009, demartino2010,demartino2013}. The light curve below
10 keV shows peculiar dips and flares on time scales of few hundreds
of seconds, suggesting an accretion nature of the X-ray emission
\citep{saitou2009, demartino2010,demartino2013}. Flares are followed
by dips in which a spectral hardening is observed, suggestive of
additional absorption by a flow of cool matter. Dips with little to
none spectral evolution also occur randomly during the quiescent
emission, and are possibly interpreted in terms of occultation by
discrete blobs of material.


The emission of the IR/optical/UV counterpart is compatible with the
sum of the thermal emission of a K2-K5 V star at a distance of
2.3--3.6 kpc, and of a hotter thermal emission coming from a surface
of larger size, compatible with an accretion disc
\citep{demartino2013}. Dips and flares take place almost
simultaneously in the UV and X-ray band; together with the observed
relative amplitudes of the flares in these two bands, this strongly
indicates that the UV emission originate from reprocessing of the
X-emission in a larger region than where the higher energy emission is
generated \citep{demartino2013}.  The presence of material around the
compact object is further indicated by the detection of several
emission lines typical of accreting systems, such as H$_{\alpha}$,
$H_{\beta}$ and HeII, \citep{masetti2006,pretorius2009}, as well as by
an optical spectrum recently obtained by NTT during March 2012 (De
Martino 2013, priv. comm.).


{\src} is positionally coincident with a moderately bright gamma-ray
source detected by {\it Fermi}-LAT, {\gsrc} \citep{demartino2010,
  hill2011,demartino2013}.  {\it Fermi}-LAT observations performed
between August 2008 and September 2010 revealed a source with spectrum
described by a power law with index $2.21\pm0.09$, cut off at
$\Gamma_{\gamma}=4.1\pm1.3$ GeV, and with a luminosity of
$L_{\gamma}=(2.3\pm0.3)\times^{34}\,d_2^2$ erg s$^{-1}\simeq L_X$
above 100 MeV \citep{hill2011}\footnote{Here, only statistical errors
  are quoted. Systematic errors can be larger by a factor $\sim$
  three. See \citealt{hill2011} for a discussion of this
  issue.}. These authors also discussed the possible association of
{\gsrc} with two radio sources detected by ATCA at 5.5 and 9 GHz,
falling in its error circle. They identified the radio counterpart of
{\src} (a faint source with a $F_{\nu}\propto\nu^{-\alpha}$ power-law
spectrum, with $\alpha=0.5\pm0.6$) as the least improbable
association, given the extremely low radio-to-gamma ray luminosity
ratio shown by the other one, most probably an AGN. No significant
variation of the gamma-ray emission of the source were found by
\citet{demartino2013}, who extended the analysis including data taken
until April 2012. Their analysis also proved that the gamma-ray
emission is concurrent with observations performed at soft
(XMM-Newton, Suzaku, Swift/XRT) and hard (RXTE) X-rays.

\section{INTEGRAL observations of {\src}}
\label{sec:igr}

\citet{demartino2010} reported the analysis of {\it INTEGRAL}/ISGRI
observations of {\src} performed between March 2003 and October 2007,
and used them together with RXTE and XMM-Newton observations to build
a 0.2-100 keV spectrum which was successfully modelled by a power-law
with index $\Gamma_X=1.70\pm0.02$.  In order to study the long-term
variability of the hard X-ray emission of {\src}, and to analyse
INTEGRAL observations simultaneous to those {\it Fermi}-LAT observations
reported by \citet[][]{hill2011} and performed between August 2008 and
September 2009, we analysed all {\it INTEGRAL} \citep{winkler2003}
observations performed from March 2003 to July 2012.

Observations performed by {\it INTEGRAL} are carried out in individual
Science Window (ScW), which have a typical time duration of about 2
ks. Here, we consider all public \is and \je observations during which
{\src} has offset angle less than 14$^o$ and 5$^o$, respectively,
adding up to a total effective exposure time of 553.7 ks for \is and
39.1 ks for \je (22.9 ks from JEM-X 1 and 16.2 ks from JEM-X 2).  Data
reduction was performed using the standard ISDC offline scientific
analysis software version 10.0.

 \begin{figure}
 \includegraphics[angle=0,width=8.6cm]{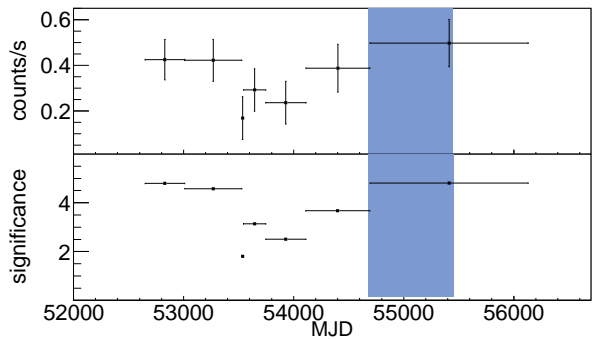}
  \caption{Long--term light curve (upper panel) and significance (lower
    panel) of {\src} on ScW timescales as seen by \is in the 18--60
    keV band. The interval covered by {\it {\it Fermi}-LAT} data
    reported by \citet{hill2011} is highlighted in blue. Here
      labels should be printed with a larger font size. \label{fig:intlc}}
  \end{figure}

While {\src} was not detected by \je at a significance larger than
3$\sigma$, we confirm the previous {\rm INTEGRAL}/ISGRI detection
reported by \citet{demartino2010}.  Combining all the ISGRI data,
{\src} is detected at a significance level of 10 $\sigma$ in the
18--60 keV band, with an average count rate of 0.365 $\pm$ 0.036 s$^{
  -1}$. Since {\src} itself is relatively faint in hard X-rays
comparing to other sources in this region, the energy spectrum was
obtained from the mosaic images. Its average spectrum can be described
by a power law with index of $1.67\pm 0.27 $, for a luminosity of
$(8.8 \pm 0.1 ) \times 10^{33}$ d$_2^2$ erg s $^{-1}$ in the 18--60 keV
band, compatible with the value derived by \citet{demartino2010} on a
smaller data set. The reduced $\chi^2$ for the fit is 0.6 under 3 d.o.f.

 To study the long-term variability of the emission observed by ISGRI
 we divided the whole exposure into seven time intervals, each with an
 exposure of roughly 80 ks. The latter is the exposure --concurrent
 with {\it Fermi}-LAT observations-- during which the source is found
 with a significance of 4.5$\sigma$.  The light curve, significance and
 effective exposure for each time span are shown in
 Figure~\ref{fig:intlc} and Table~\ref{table1}. The source is detected
 at a significance $\ga 3$-$\sigma$ in all but the third interval,
 which is the one covering the shorter time period (15 days), and
 during which the significance falls to $1.8$-$\sigma$. However,
 modelling the overall light curve with a constant gives a $\chi^2=9$ over six
 degrees of freedom, clearly indicating that the observed emission is
 compatible with a constant.

\begin{table}
\centering
\caption{Flux, detected significance and effective exposure time of 7 evenly divided time spans of ISGRI observations of {\src}}             
\label{table1}      
\centering                          
\begin{tabular}{cccccccccc}        
\hline\hline                 
Time covered (MJD)  & Intensity (s$^{-1}$) & Signif. ($\sigma$)  & Expos. (ks)\\
  \\
\hline
52650.7 -- 53010.2&  0.42&  4.79&  79.2\\
53010.2 -- 53528.3&  0.42&  4.58&  78.9\\
53528.4 -- 53543.5&  0.17&  1.81&  80.1\\
53543.6 -- 53746.9&  0.29&  3.14&  79.3\\
53746.9 -- 54110.9&  0.24&  2.50&  79.2\\
54111.1 -- 54692.7&  0.39&  3.67&  79.6\\
54693.7 -- 56131.0&  0.50&  4.80&  79.1\\

\hline
\hline                                   
\end{tabular}
\end{table}

Finally, in order to search for any periodic signal in the long--term
light curve, we use the Lomb--Scargle periodogram method
\citep{lomb1976,scargle1982}. Power spectra are generated for the
light curve using the PERIOD subroutine \citep{press1989}. The 99\%
white noise significance levels are estimated using Monte Carlo
simulations \citep[see e.g.][]{kong1998}. No signal was significantly
detected beyond such noise level. 

Our analysis indicates that {\src} keeps behaving as a steady hard
X-ray emitter, over a 9-year time interval. Also, it shows that the
source was active in hard X-rays simultaneously to the Fermi
observations performed between August 2008 and September 2010 and
analysed by \citet{hill2011}, confirming the simultaneous RXTE/Fermi
detection achieved during 2011 \citep{demartino2013}.

\section{Propeller state}
\label{sec:propeller}

The fate of matter in-falling towards a magnetised rotating neutron
star depends essentially on the ratio between the rotation rate of the
in-flowing matter and of the magnetosphere, evaluated at the radius
where the dynamics of the flow becomes dominated by the stress
exerted by the magnetic field, the so-called disk truncation radius
$R_{\rm in}$ \citep[see, e.g.][and references therein]{ghosh2007}.  In
a Keplerian this disk, matter rotates at a rate:
\begin{equation}
\Omega_{\rm K}(r)=(GM_*/r^3)^{1/2},
\end{equation}
where $M_*$ is the mass of the compact object, and it is useful to
define the fastness parameter as the ratio between the neutron star
rotation rate, $\Omega_*=2\pi/P$, and the Keplerian rate at the truncation
radius:
\begin{equation}
\label{eq:fastness}
\omega_*=\frac{\Omega_*}{\Omega_{\rm K}(R_{\rm in})}=\left(\frac{R_{\rm in}}{R_{\rm c}}\right)^{3/2}.
\end{equation}
Here
\begin{equation}
\label{eq:rcor}
R_{\rm c}=(GM_*/\Omega_*^2)^{1/3}
\end{equation} is the co-rotation radius.
While for $\omega_*<1$ ({\it slow} rotator case), mass in the disc is
allowed to accrete yielding its specific angular momentum to the
neutron star, for $\omega_*\geq1$ ({\it fast} rotator case), the
inflowing mass finds a centrifugal barrier which partly or completely
inhibit further in-fall onto the surface of the neutron star.

Such a bi-modality of the outcome of accretion follows from the nature
of the coupling between the field lines and the disc matter. In order
to flow towards the compact object in a Keplerian disk at a steady
rate $\dot{M}$, matter has to get rid of its angular momentum at a
rate:
\begin{equation}
\label{eq:amloss}
(d/dt)L_{\dot{M}}(r)=\dot{M}\Omega_{\rm K}(r) r^2.
\end{equation}
Far from the central object, it is disk viscosity which redistributes
this angular momentum towards the outer rings of the disk. As matter
approaches a magnetised neutron star, the stress exerted by its
rotating magnetic field becomes dominant. Differential rotation
between the field lines, assumed to be initially poloidal, and disk
matter yields a stress:
\begin{equation}
S^{\rm m}=\pm B_pB_\phi/4\pi,
\end{equation}
where $B_\phi$ is the toroidal component of the field originated from
the twisting of the poloidal component 
\begin{equation}
B_p(r)=\mu r^{-3},
\end{equation} and $\mu$
is the NS magnetic dipole moment. Reconnection and opening of the
field lines limit the magnitude of the toroidal component to a
fraction $\eta\la1$ of the poloidal component, and reduce the
interaction layer to a width $(\Delta r/R_{\rm in})<1$
\citep[e.g.][]{wang1996,ghosh2007,lovelace1995,romanova2009,dangelo2010}. The
magnetic torque integrated on such layer can be thus written as
\citep[see, e.g.,][]{dangelo2010}:
\begin{eqnarray}
\label{eq:magtorque}
T_{\rm m}&=&2\int [r S^{\rm m}(r)] r dr d\phi=4\pi
S^{\rm m}(R_{\rm in}) R_{\rm in}^2 \Delta r=\nonumber\\ &=&\pm \eta
\left(\frac{\Delta r}{R_{\rm in}}\right) \frac{\mu^2}{R_{\rm in}^3},
\end{eqnarray}
where $[r S^{\rm m}(r)]$ is the torque acting per unit area, and the
factor 2 reflects the faces of the disc over which the torque is
applied.

The sign of the magnetic torque exerted by the NS on the disk matter
(Eq.~\ref{eq:magtorque}) depends on the direction of the twist, and is
positive when $\omega_*>1$. In such conditions the neutron star
deposits angular momentum into the disc, which makes feasible the
ejection of matter along the field lines \citep[i.e., a propeller
  state;][]{illarionov1975}.

However, for values of $\omega_*$ between $1$ and a certain critical
value $\omega_*^{\rm cr} \ga 1$, the energy released by the
propellering magnetosphere to the disc plasma is not sufficient to
unbind it from the system \citep{spruit1993}. A large fraction of the
propellered matter returns to the disc and builds up there, and may
eventually resume accretion as it increases the inward accretion rate
\citep[see the recent works by][who examined in depth the cycles
  between accretion and angular momentum deposition at the inner rim,
  which takes place for values of $1<\omega_*\leq\omega_{\rm
    cr}$]{dangelo2010,dangelo2012}. For values of the fastness
$\omega_* > \omega_*^{\rm cr}$ the ejection of matter at the inner rim of the disc
is instead clearly favoured, as the angular momentum and the energy
which may be released by the NS to the disc matter increases. This
tendency is also confirmed by the magneto-hydrodynamical simulations
performed by \citealt{romanova2003,ustyugova2006,romanova2009}. In the
following, we consider a similar situation to model the phenomenology
shown by {\src}, and assume for simplicity that all the incoming
matter is ejected by the system, $\dot{M}_{\rm ej}=\dot{M}$. 

\begin{figure}
 \includegraphics[angle=0,width=\columnwidth]{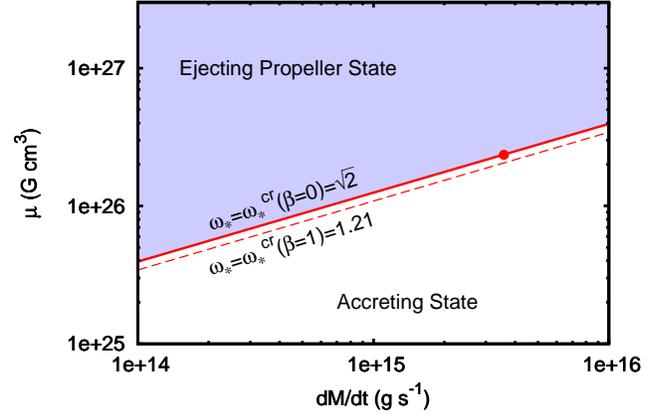}
  \caption{Values of the NS dipole moment and of the rate of mass lost
    by the disc giving the critical values of the fastness parameters,
    for a 1.4 M$_{\odot}$ NS spinning a a period of 2.5 ms, with
    $\eta=1$ and $(\Delta r/R_{\rm in })=0.1$. The red solid and
    dashed lines are evaluated for the critical fastness of the purely
    an-elastic and elastic case, respectively. These lines delimit the
    blue shaded region, where fully ejecting propeller solutions
    hold. The red circle marks the mass ejection rate evaluated for a
    system with a spin period of 2.5 ms, a luminosity of $10^{35}$ erg
    s$^{-1}$, and a fastness equal to the critical value.
 \label{fig:mumdot}}
  \end{figure}

Under the above assumptions, the conservation of angular momentum at
the inner rim of the disc reads as:
\begin{eqnarray}
\label{eq:ejection} 
\dot{M} R_{\rm in} v_{\rm out}&=&(d/dt){L}_{\dot{M}}+T_{\rm m} = \nonumber \\ 
&=&\dot{M}\Omega_{\rm K}R_{\rm  in}^2+\eta\left(\frac{\Delta r}{R_{\rm in}}\right)\frac{\mu^2}{R_{\rm in}^3},
\end{eqnarray} 
where, the term on the left hand side is the rate of angular momentum
lost in the outflow, while the rate of angular momentum carried by
disk matter (Eq.~\ref{eq:amloss}) and the torque applied by the
magnetic field lines (Eq.~\ref{eq:magtorque}) appear in the right hand
side. \citet{eksi2005} proposed an useful parametrisation of the
propeller process in terms of the elasticity of the scattering between
the field lines and the disc plasma through a parameter $\beta$, which
varies between $\beta=1$ in the perfectly elastic case, and $\beta=0$
in the purely an-elastic case. According to this parametrisation, the
velocity of the outflow is:
\begin{equation}
\label{eq:veloutflow}
v_{\rm out}= \Omega_{\rm K}(R_{\rm in}) R_{\rm in} [1-(1+\beta)(1-\omega_*)].
\end{equation}
Substituting this relation in Eq.\ref{eq:ejection} yields:
\begin{equation}
\label{eq:fullpropomg}
\omega_*^{7/3}(\omega_*-1)(1+\beta)=\frac{\eta(\Delta r /
  R_{\rm in})\mu^2}{\dot{M}\sqrt{GM}R_{\rm c}^{7/2}}.
\end{equation} 
An ejecting propeller solution may hold only if the fastness exceeds
the critical threshold, which was estimated by \citet{perna2006} as:
\begin{equation}
\label{eq:omgcr}
\omega_*^{\rm cr}(\beta)=\frac{\beta+\sqrt{2}}{1+\beta}.
\end{equation}
The critical threshold takes a value of $1.21$ and $\sqrt{2}$ for the
perfectly elastic ($\beta=1$) and an-elastic ($\beta=0$) case,
respectively. We plot in Fig.~\ref{fig:mumdot} the values of the NS
dipole moment and rate of mass lost leading to such critical values,
which delimit the region where fully ejecting propeller solutions are
possible.

The energy available to power the observed emission follows from the
conservation of energy \citep{eksi2005}:
\begin{eqnarray}
\label{eq:mdot}
L_{\rm rad}&=&\frac{GM\dot{M}}{R_{\rm in}}+\Omega_* T_{\rm
  m}-\frac{1}{2}\dot{M}v_{\rm out}^2=\\
&=&
 \frac{GM\dot{M}}{2R_{\rm
    in}}[1+(1-\beta^2)(\omega_*-1)^2],
\end{eqnarray}
and is related to the rate of mass in-flow (which equals the mass
ejection rate under the assumptions made). Considering the critical
fastness for the an-elastic case ($\omega_*^{\rm cr}=\sqrt{2}$,
$\beta=0$), a spin period of 2.5 ms, and a value of the luminosity of
$10^{35}$ erg s$^{-1}$ (see Sec.~\ref{sec:model}), gives a mass
ejection rate of $3.6\times10^{15}$ g s$^{-1}$, which crosses the
relative propeller solution for $\mu\ga\mbox{few}\times10^{26}$ G
cm$^3$ (see the red circle in Fig.~\ref{fig:mumdot}), of the order of
the field usually estimated for NS in a LMXB.

 In Sec.~\ref{sec:sed}, we interpret the X-ray and $gamma$-ray
  emission of {\src} in terms of the synchrotron and self-synchrotron
  Compton emission by a population of relativistic electrons,
  accelerated by shocks at the magnetospheric interface of a
  propellering neutron star, and interacting with the NS magnetic
  field at the interface, $\bar{B}$. To estimate the strength of such
  field we consider a value of the order of the dipolar component of
  the NS field, evaluated at the inner disc radius:
\begin{eqnarray} 
\label{eq:dipole}
\bar{B}& =& B_p(R_{\rm in})=\frac{\mu}{R_{\rm in}^{3}}  = \frac{\mu}{R_{\rm c}^{3}\omega_*^{2}} = \frac{\mu\Omega_*^2}{GM_*\omega_*^2}=\frac{4\pi^2\mu}{GM_*P^2\omega_*^2}=\nonumber \\
&=&0.85\times10^6\;\mu_{26}\;P_{2.5}^{-2}\;(\omega_*/2)^{-2}\;G,
\end{eqnarray}
where $\mu_{26}$ is the NS magnetic dipole moment in units of
$10^{26}\, \mbox{G cm}^{3}$, and $P_{2.5}$ the spin period in units of 2.5
ms. The expression in the right hand side is obtained by using the
definition of the NS fastness (Eq.~\ref{eq:fastness}), and the
definition of the corotation radius (Eq.~\ref{eq:rcor}), considering a
NS mass of 1.4 M$_{\odot}$ (as implicitely assumed in the rest of the
paper), taking a screening coefficient $\eta=1$, and ignoring the
tangential component introduced by shearing. The Eq.~\ref{eq:dipole}
implicitely expresses the NS magnetic dipole moment in terms of the
field strength at the interface, the NS spin period, and the fastness,
and can be plugged in the expression of the angular momentum
conservation, Eq.~\ref{eq:fullpropomg}. Further, $\dot{M}$ can be
expressed as a function of the radiated luminosity, spin period,
fastness and elasticity parameter thanks to the relation expressing
energy conservation, Eq.~\ref{eq:mdot}. Substituting in
Eq.~\ref{eq:fullpropomg}, and setting $\eta=1$ and $(\Delta r/R_{\rm
  in})=1$, finally yields:
\begin{equation}
\label{eq:solution}
\bar{B}=5.2\times10^6\;L_{35}^{-1}\;P_{2.5}^{-1/2}\left[\frac{1}{\omega_*}\frac{(\omega_*-1)\times(1+\beta)}{1+(1-\beta^2)(\omega_*-1)^2}\right]^{1/2}.
\end{equation}
 Considering a spin period in a range typical of millisecond pulsars
 (1.5--5 ms), and a value of the fastness exceeding the critical
 threshold for mass ejection, $\omega_*^{\rm cr}$
 (Eq.~\ref{eq:omgcr}), we conclude that the magnetic field at the
 interface $\bar{B}$ must lie in a range between $2.2$ and
 $11\times10^6$ G, to produce a total propeller luminosity of
 $1.5\times10^{35}$ erg s$^{-1}$ (see Sec.~\ref{sec:results} and
 Table~\ref{table}).

It has to be noted that mass ejection is not a necessary outcome of a
system in a propeller state. In fact, a steady solution for a thin
disc with an angular momentum source at the inner rim exists for all
values of $\omega_*>1$ \citep{syunyaev1977}. In such case, the angular
momentum is retained in the disc, which readjusts by increasing its
density with respect to the standard accreting solution, in order to
match the increased demand of viscosity set by the source of angular
momentum at the inner boundary.  No mass in-flows in this case and the
disc is considered {\it dead}. The angular momentum injected by the NS
at the inner rim is released at the outer edge of the disc, most
probably to the orbit of the binary through tidal interactions. Such
state has been recently re-examined by
\citet{dangelo2011,dangelo2012}. However, while a dead disc solution
exists even if no matter is ejected by the system, it hardly holds on
year-long time-scales such as those observed in {\src}. As a matter of
fact, if the disc is continuously replenished by a source of mass at a
rate $\geq ~10^{-12}$ M$_{\odot}$ yr$^{-1}$, like those commonly
observed from LMXB \citep[e.g.][]{coriat2012}, it takes the inward
pressure of a {\it dead} disc only a few months to bring the inner
disk radius back to the co-rotation surface. We then consider only the
full-ejecting case discussed above as a possibility to explain the
properties of {\src} in terms of a propeller state.


\section{Spectral energy distribution}
\label{sec:sed}

The plasma of the layer where the accretion disc is truncated in a
propeller state is expected to be very turbulent and magnetised, as a
result of the deposition of a copious amount of energy by the
magnetosphere (see, e.g., magnetohydrodynamics numerical simulations
studied by \citealt{romanova2009} and references therein).  Such
region was identified by \citet{bednarek2009,bednarek2009b} as a
suitable site to accelerate charged particles to relativistic energies
through a Fermi process.  Here, we apply a similar guess to the case
of a relatively weakly magnetised ($B_{\rm NS}\approx10^{8-9}$ G, $\mu
\approx 10^{26}$--$10^{27}$ G cm$^{3}$), quickly spinning ($P \approx
\mbox{few}$ ms) NS in a LMXB, and study the spectral energy
distribution expected to arise from the population of relativistic
electrons expected to be produced at the layer between the
magnetosphere and the disc.  For simplicity, we consider in the
following that the electron distribution occupies a torus-like volume,
with radial size equal to the inner disc radius $R_{\rm in}=R_{\rm
  c}\omega_*^{2/3}$ (see Eq.~\ref{eq:fastness}), and transverse
section of size, $R_{\rm t}$. Only the acceleration of electrons is
considered in this model, while the possible contribution of hadrons
is discussed in Sec.~\ref{sec:hadrons}.

\subsection{Electrons acceleration}

In a Fermi acceleration process, energy is given up to each electron at a
rate {\setlength\arraycolsep{0.1em}
\begin{eqnarray}
\label{eq:engain}
\ell_{acc}&=&\xi c E / R_L=\xi e c B(R_{\rm in})=1.4\times10^5 \xi_{0.01}\;\bar{B}_6\:\mbox{erg s}^{-1},
\end{eqnarray}
where $R_L=E/e B $ is the Larmor radius, $\xi_{0.01}$ is the
acceleration parameter in units of 0.01, $e$ is the electron charge,
and $\bar{B}_6$ is the strength of the magnetic field at the interface
$\bar{B}$, in units of $10^6$ G, which is of the order of the
  values determined in Sec.~\ref{sec:propeller}.

The time scale of electron acceleration is:
\begin{equation}
\label{eq:tauacc}
\tau_{\rm acc}=\frac{\gamma m_e c^2}{\ell_{\rm acc}}=5.7\times10^{-8}\;\xi_{0.01}\;\bar{B}_6\;(\gamma/10^4)\; \mbox{s},
\end{equation}
where $m_e$ is the electron mass. This value is much shorter than
the time needed to travel the typical size of the region, $R_{in}$,
\begin{eqnarray}
\label{eq:tautravel}
\tau_{\rm tr}&=&
\frac{R_{\rm in}}{c}=\frac{R_{\rm c}\omega_*^{2/3}}{c} = \nonumber \\
& \approx & 1.6\times10^{-4}\;P_{2.5}^{2/3}\;(\omega_*/2)^{2/3}\;\mbox{s},
\end{eqnarray} 
ensuring that  electrons can be effectively accelerated before that they can escape the system.

\subsection{Emission processes and expected dominant components}

The electrons accelerated by a Fermi process lose energy through the
emission of radiation produced by their interaction with the magnetic
and the radiation field permeating the transition layer, and with the
ions of the plasma.

\subsubsection{Synchrotron  emission}
\label{sec:synchro}
Synchrotron losses proceed in the transition layer at the rate set by the Larmor formula:
\begin{eqnarray}
\label{eq:synchro}
\ell_{syn}=\frac{4}{9}\frac{e^4}{m_e^2 c^3} \bar{B}^2 \gamma^2 
 =1.1\times10^5\;\bar{B}_6^2\;(\gamma/10^4)^2\;\mbox{erg
  s}^{-1},
\end{eqnarray}
where $\gamma$ is the electron Lorentz factor.  If synchrotron losses are dominant over other
channels of energy losses (see below), the parameters describing the
electron energy distribution are set by the equilibrium between the
energy injection through Fermi acceleration and synchrotron
emission. In particular, the cut-off energy of the electron
distribution is set by equating Eq.~\ref{eq:engain} and
\ref{eq:synchro}:
\begin{equation}
\label{eq:gammamaxsyn}
\gamma_{\rm max}^{syn}=\frac{3}{2}\frac{m_e
  c^2}{e^{3/2}}\left(\frac{\xi}{\bar{B}}\right)^{1/2}=1.2\times10^4\;\xi_{0.01}^{1/2}\;\bar{B}_6^{-1/2}.
\end{equation} Assuming an exponentially cut power law distribution for the
energy of the electrons:
\begin{equation}
\label{eq:elpopul}
\frac{dN_e}{d\gamma}=K \gamma^{-\alpha}\exp{\left(-\frac{\gamma}{\gamma_{\rm max}}\right)},
\end{equation}
the synchrotron spectral energy distribution is described by a power
law: 
\begin{equation}
(E F_E)^{syn}\propto E^{-(\alpha-3)/2}\exp{\left[-\frac{3}{2}\left(\frac{E}{E_{\rm max}^{syn}}\right)^{1/3}\right]},
\end{equation} with cut-off energy:
\begin{equation}
\label{eq:synen}
E_{\rm max}^{syn}=\frac{3}{2}\frac{\hbar}{m_e c} \bar{B} (\gamma_{\rm max}^{syn})^2=\frac{27}{16} \frac{\hbar m_e c^3}{e^2}\xi=1.2 \;\xi_{0.01}\;\mbox{MeV}.
\end{equation} 
\citep[see, e.g., the relations given by][evaluated for an electron
  distribution like the one given by Eq.~\ref{eq:elpopul}, and setting
  the electron Lorentz factor to the value given by
  Eq.~\ref{eq:gammamaxsyn}]{lefa2012}. It results that when the
synchrotron emission is the dominant cooling process, the cut-off
energy of the emitted spectrum depends only on the acceleration
parameter, $\xi$. This parameter may take values $<<1$ in the case of
relativistic shocks, but it is largely undetermined on theoretical
grounds \citep[see][and references therein]{khangulyan2007}. In our
model we assume that the main contribution to the 0.2--100 keV
spectrum of \src\ (a power law $E F_{E}\propto E^{-(\Gamma_X-2)}$,
with an index $\Gamma_X=1.70\pm0.02$ and no cut-off detected up to 100
keV, see Sec.~\ref{sec:xss} and \ref{sec:igr}) is given by synchrotron
emission. Imposing that the cut-off energy of this component lies
between 100 keV and 100 MeV, Eq.~\ref{eq:synen} can be used to
constrain $\xi$ to a broad range, $8.5\times10^{-4}$--0.85. Similarly
the observed spectral slope indicates an electron energy distribution
with $\alpha\simeq2\Gamma-1=2.4$.  On the other hand, the high energy
part of the spectrum observed by {\it Fermi}-LAT, with a cut-off at
$4.1\pm1.3$ GeV \citep{hill2011}, cannot be explained by synchrotron
emission alone, as it would require $\xi\sim30$, and is instead
discussed in terms of (self-synchrotron) inverse Compton emission in
Sec.~\ref{sec:iccem}.

At low energies the emitting region becomes optically thick to the
synchrotron radiation. We evaluate the absorption coefficient for the
relevant parameters of the system and an electron distribution with
index $\alpha=2.4$, following \citet{rybicki1979}:
\begin{equation}
\alpha_{syn}(E)=3\times10^{-4}\;\left(\frac{n_e}{10^{17}\mbox{cm}^{-3}}\right)\;\bar{B}_6^{2.2}\;\left(\frac{E}{\mbox{eV}}\right)^{-3.2}\;\mbox{cm}^{-1},
\end{equation}
where $n_e$ is the density of electrons of the considered medium,
scaled to a value of the order of those obtained through modelling of
the observed spectrum (see below). Imposing $\tau=\alpha_{syn}(E_{\rm br})R_{\rm
  t}=1$, we estimate the energy below which the medium becomes
optically thick to synchrotron radiation as:
\begin{equation}
\label{eq:lowenbreak}
E_{\rm br}=2.9\;\left(\frac{n_e}{10^{17}\mbox{cm}^{-3}}\right)^{0.31}\;\left(\frac{R_t}{\mbox{km}}\right)^{0.31}\;\bar{B}_6^{0.69}\;\;\mbox{eV}.
\end{equation} 
This value is between the optical and the UV band for typical
parameters of the system, compatible with the absence of a low-energy
cut-off in the observed X-ray data.

\subsubsection{Inverse Compton emission}
\label{sec:iccem}

The weight of inverse Compton losses in the Thomson domain with respect to synchrotron
losses can be evaluated as:
\begin{equation}
\label{eq:ratio}
\frac{\ell_{IC}}{\ell_{syn}}=\frac{\epsilon_{ph}}{\epsilon_{\rm mag}},
\end{equation}
where $\epsilon_{\rm ph}$ and $\epsilon_{\rm mag}=\bar{B}^2/8\pi$ are the
energy density in the radiation and in the magnetic field,
respectively. 

The inner rings of a viscous disc emit thermal photons with typical
temperature set by amount of angular momentum that has to be
dissipated by the disc. This is related to the rate of mass inflow
(here set equal to the rate of mass lost) and by the size of the disc
\citep{frankkingraine2002}, yielding $kT(R_{\rm in})\approx 100$ eV
for $\dot{m}_{15}=P_{2.5}=1$ and $\omega_*=2$.  Even ignoring the
reduction of the cross section for inverse Compton scattering of
photons with initial energy lower than the Klein-Nishina threshold
$\approx m_e c^2 / 4 \gamma$ ($\approx 10$ eV for $\gamma=10^4$), the
energy density implied by a similar thermal spectrum,
$\epsilon_{disc}=aT^4(R_{\rm in})$, is lower by more than a factor
$1000$ than the density associated to a $10^6$ G magnetic
field. Inverse Compton scattering of disc photons is then
energetically unimportant with respect to synchrotron emission, for
typical parameters of the systems considered here,

Photons emitted by the low mass companion have an even  lower
density at the interface between the disc and the magnetosphere,
$\epsilon_{star}/\epsilon_{disc}\approx[T_2/T(R_{\rm
    in})]^4(R_2/a)^2\approx10^{-10}$, where $T_2$ and $R_2$ are the
companion star temperature and radius, respectively, and $a$ is the
size of the orbit. To evaluate such ratio we considered
$T_2\simeq4600\,K$ , $R_2\approx 0.6 R_{\odot}$, a total mass of the
system of 2 M$_{\odot}$, and an orbital period of 8 hr, values typical
of a late type K star as proposed by \citet{demartino2013}.

On the other hand, inverse Compton scattering of the synchrotron
photons off the same electron population which produced them
(synchrotron self Compton process; SSC in the following) may play an
important role if the electron distribution is concentrated in a
relatively small region, such as the transition region that we are
considering here ($R_{\rm in}\approx 50$ km, $R_{\rm t}\approx
\mbox{few km}$).  In the Thomson regime, the ratio between the
luminosity emitted through synchrotron and SSC process is
\citep[e.g.][]{sari2001}:
\begin{equation}
\label{eq:SSCratio}
\frac{L_{\rm SSC}}{L_{\rm syn}}\sim \frac{1}{3}\frac{E_{\rm
    max}^{SSC}}{E_{\rm max}^{syn}}\sigma_T n_e R_{\rm in},
\end{equation} 
where $E_{\rm max}^{SSC}$ and $E_{\rm max}^{syn}$ are the peak
energies of the SSC and synchrotron photons. While the latter energy
is set by Eq.~\ref{eq:synen}, the cut-off of the inverse Compton
distribution reproduces that of the electron energy distribution,
$E_{\rm SCC}^{max}=\gamma_{max}m_ec^2$.  Taking into account also the
SSC losses, and defining $f\equiv 1+ L_{\rm SSC}/L_{\rm syn}$, the
maximum energy that can be achieved by the electron distribution is
then obtained by balancing electron energy losses and gains,
\begin{equation} 
\label{eq:balance}
\ell_{syn}+\ell_{SSC}=f\ell_{syn}=\ell_{acc},
\end{equation} yielding:
\begin{equation}
\label{eq:gammamax}
\gamma_{\rm max}^{SSC}=\frac{\gamma_{\rm max}^{syn}}{\sqrt{f}}=8.2\times10^3\;f_{2}^{-1/2}\:\xi_{0.01}^{1/2}\;\bar{B}_6^{-1/2}.
\end{equation} 
Here, we defined $f_2=f/2$, to scale the energy ratio to the case of
an equal luminosity released by synchrotron and SSC processes. The SSC
spectrum is then cut-off at an energy of:
\begin{equation}
\label{eq:enmax}
E_{\rm SCC}^{max}=4.2\;f_{2}^{-1/2}\:\xi_{0.01}^{1/2}\;\bar{B}_6^{-1/2}\; \mbox{GeV}.
\end{equation}
of the order of that observed from {\src} by {\it Fermi}-LAT
($4.1\pm1.3$ GeV; \citealt{hill2011}). SSC emission can be thus
responsible of the $\gamma$-ray flux observed from {\src} and perhaps
from other binaries. By setting the high-energy cut-off of the
spectrum to the observed value, and varying the magnetic field in the
range determined in Sec.~\ref{sec:propeller} ($\bar{B}_6$=2.2--11),
Eq.~\ref{eq:enmax} shows that the ratio of the luminosity emitted by
SSC and synchrotron process depends linearly on the poorly constrained
acceleration parameter, $\xi$. If the latter is varied in the range
determined in Sec.~\ref{sec:synchro} by imposing that the cut-off of
the synchrotron spectrum lies between 100 keV and 100 MeV
($\xi=8.5\times10^{-4}$--0.85), values of $f$ ranging from
$\sim10^{-2}$ to 40 are obtained. It is then clear that a sensible
estimate of the expected flux ratio cannot be given without an
accurate knowledge of the acceleration parameter. On the other hand,
as {\src} emits a comparable $\gamma$-ray and X-ray flux, we expect
the two components to emit energy at a comparable rate.  By setting
$f=2$ in Eq.~\ref{eq:enmax}, we can therefore constrain the
acceleration parameter to lie in the range $\xi=0.02$--$0.10$, in
order to reproduce the observational features of {\src}, in the
framework set by our model.

At zero order, the electron density requested to produce a
  similar contribution from SSC and synchrotron emission ($L_{\rm
    SSC}/L_{\rm syn}=1$, $f=2$) can be estimated from Eq.~\ref{eq:SSCratio}:
\begin{equation}
\label{eq:eldens}
n_e\approx2.7\times10^{14}\;f_{2}^{-1/2}\:\xi_{0.01}\;P_{2.5}^{-2/3}\;(\omega_*/2)^{-2/3}\;\mbox{cm}^{-3}.
\end{equation}
However, this value is largely underestimated as the reduction of the
cross section due to Klein Nishina effects largely decreases the
efficiency of the SSC emission. We show in Sec.~\ref{sec:model} how
densities of the order of $10^{17}\,\mbox{cm}^{-3}$ are needed to
reproduce the observed spectral energy distribution.

This value of the electron density has to be compared with the
expected density in the outflow. By imposing mass continuity at the
base of the outflow, we estimate a scale unit of the mass density as:
\begin{eqnarray}
\rho_0&=&\frac{\dot{M}}{2\pi R_{\rm in} R_{\rm t} v_{\rm
    out}}\nonumber=\\&=&10^{-7}\;\dot{m}_{15}\;P_{2.5}^{-1/3}\;(\omega_*/2)^{-4/3}\;(R_{\rm
  t}/10^5\mbox{cm})\; \mbox{g cm}^{-3},
\end{eqnarray}
where we used Eq.~\ref{eq:veloutflow} to express the outflow velocity
in the purely ejecting case ($\beta=1$), and considered a typical size
for the transverse section of the acceleration region $R_{\rm
  t}\approx 10^5$ cm (see below). For a fully ionised plasma, the
electron density is therefore:
\begin{equation}
n_{e,0}\simeq\frac{\rho_0}{m_{H}}(X+Y/2)\approx0.5\times10^{17}\;\mbox{cm}^{-3},
\end{equation}
where we omitted the dependencies on the scale units used so far, and
we considered solar abundances for hydrogen and helium, $X=0.7$ and
$Y=0.28$, respectively. Such a value of the scale unit of electron density is not far from that needed to produce a comparable SSC and synchrotron
emission ($\approx\mbox{few}\times10^{17}\,\mbox{cm}^{-3}$; see Sec.~\ref{sec:model}).

\subsubsection{Bremsstrahlung emission}

To evaluate the energy lost by electrons in bremsstrahlung
interactions with the ions of plasma, we considered the relation given
by \citet{blumenthal1970}, in the approximation of a fully ionised plasma:
\begin{eqnarray}
\ell_{brems}&=&\frac{4e^6}{\hbar m_e^2 c^4}
(2n_{H}+20n_{He})[\log(2\gamma)-1/3]\gamma m_e c^2\simeq\nonumber\\
&&0.66\;\left(\frac{\rho}{\rho_0}\right)\;\left(\frac{\gamma}{10^4}\right)\;
\mbox{erg s}^{-1}.
\end{eqnarray}
Here $n_H=X \rho/m_H$ and $n_{He}=Y \rho/m_H$ are the hydrogen and
helium number densities, respectively, and a solar composition is
considered. By comparing this relation to energy gains and other
radiative losses (see Eq.~\ref{eq:engain}, \ref{eq:synchro} and
\ref{eq:balance}), we deduce that bremsstrahlung radiation is not
dominant for the typical densities of a propeller outflow.

\subsubsection{Coulomb losses}

Energy losses through Coulomb interactions can be safely ignored for
the relevant parameters.  Considering the rate of energy lost by a
relativistic population of electrons through Coulomb interactions with
other electrons and with much less energetic ions
\citep[e.g.][]{frankel1979}, a slowing down time scale:
\begin{equation}
\tau_{\rm coll}\simeq 130\;(\gamma/10^4)\;(\rho/\rho_0)^{-1}\;\mbox{s}
\end{equation} is obtained, much longer than other time scales of the system.

 \begin{figure}
 \includegraphics[angle=0,width=\columnwidth]{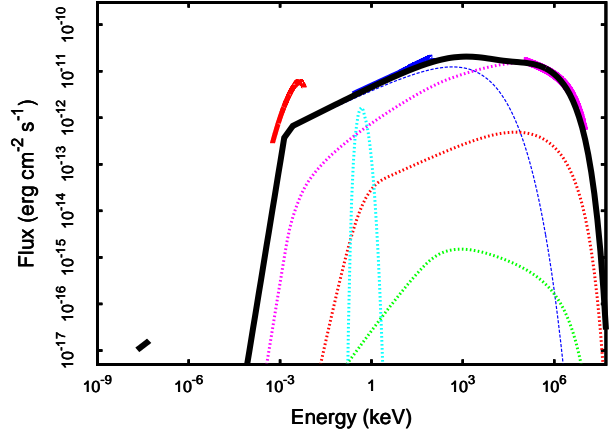}
  \caption{Spectral energy distribution observed from {\src} in the
    $\gamma$-rays, X-rays, IR/optical/UV and radio bands (magenta,
    blue, optical and black thick lines, respectively). The model
    obtained for $\alpha=2.4$, $\gamma_{\rm max}=10^4$,
    $\bar{B}=5\times10^6$ G and $n_e=12\times10^{17}$ cm$^{-3}$ is
    over-plotted as a black line. The components due to the
    synchrotron, SSC, inverse Compton of disc photons, and
    bremsstrahlung in the relativistic plasma are shown as blue,
    magenta, red and green dotted lines, respectively. The cyan dashed
    line represents the possible contribution of the inner parts of an
    accretion disk truncated at a radius of 50 km, with an inner
    temperature of 0.1 keV, an inclination of $75^{\circ}$, and
    absorbed by the interstellar medium with an absorption column
    $N_{\rm H}=10^{21}$ cm$^{-2}$ \citep{demartino2010}. This
    component has not been taken into account in evaluating the model,
    considering the large uncertainties on the actual emission of a
    disk in a propeller state.\label{fig:spectrum}}
  \end{figure}

\subsection{Simulated spectral energy distribution}
\label{sec:model}

To estimate quantitatively if the proposed model can reproduce the
broad-band spectral energy distribution observed from {\src}, we
modelled the radiation processes described in the previous section
(synchrotron, SSC, inverse Compton scattering of the disc photons, and
bremsstrahlung) using the codes described by
\citet{torres2004}, \citet{decea2009} and \citet{martin2012}, and assuming a distance to
the system of 2 kpc.

The electron distribution for $\gamma>1$ is described by an
exponentially cut-off power law (Eq.~\ref{eq:elpopul}).  The energy
distribution and luminosity produced by the synchrotron and SSC
processes also depend on the strength of the magnetic field
interacting with the electrons, $\bar{B}$, and on the electron
density, $n_e$ respectively.

\begin{table*}
\centering

 \begin{minipage}{140mm}

  \caption{Model parameters reproducing the X-ray to $\gamma$-ray
    spectrum observed from {\src}. The cut-off of the electron
    distribution is fixed at $\gamma=10^4$ and $\alpha=2.4$. A
    distance of 2 kpc is considered.\label{table}}
  \begin{tabular}{@{}cccc||ccccccc@{}}
  \hline
  \hline
$\bar{B}$ (MG) & $n_{e,17}$ (cm$^{-3}$) & $L_{\rm syn;35}$  & $L_{\rm SSC;35}$  & $\xi$ & $\omega_*$ & $\beta$ & $\mu_{26}$ (G cm$^{3}$) & $\dot{m}_{15}$ (g s$^{-1}$)& $ R_{\rm in}$ (km) & $ R_{\rm t}$ (km)\\
\hline
$1$ & $2.6$ & $0.48$ &  $0.77$ & $0.019$ & $1.03$ &{...} & $0.3$ & {...} & $31$ & $1.3$ \\
$3$ & $4.9$ & $0.62$ & $0.67$ & $0.046$ & $1.40$-$1.44$ & $0-0.1$ & $1.7-1.8$ & $4.5-4.6$ & $38-39$ & $0.3-0.4$ \\
$4$ & $7.0$ & $0.73$ & $0.71$ & $0.059$ & $1.3-1.70$ & $0.40-1$ & $1.9-3.4$ & $4.8-5.6$ & $36-44$ & $0.20-0.22$\\
$5$ & $12$ & $0.77$ &  $0.72$ & $0.072$ & $1.4-1.9$ & $0.76-1$ & $3.1-5.4$ & $5.6-6.2$ &  $39-48$ & $0.12-0.13$ \\
$6$ & $12$ & $0.78$ &  $0.69$ & $0.083$ & $1.8-2.4$ & $0.93-1$ & $5.8-10$ & $7.0-7.3$ & $46-56$ & $0.09-0.10$ \\  

\hline
\end{tabular}
\end{minipage}

\end{table*}

 We set the model parameters to reproduce the $\Gamma_1=1.70\pm0.02$ power
 law observed in the 0.2--100 keV band by {\it XMM-Newton}, {\it RXTE}
 and {\it INTEGRAL} (with an unabsorbed flux of
 $(4.5\pm0.9)\times10^{-11}$ erg cm$^{-2}$ s$^{-1}$;
 \citealt{demartino2010,demartino2013}), and the $\Gamma_2=2.21\pm0.09$ power
 law, cut off at $E_{\rm cut}=4.1\pm1.3$ GeV, observed above 100 MeV by {\it
   Fermi}-LAT (with an unabsorbed flux of $(4.1\pm0.3)\times10^{-11}$
 erg cm$^{-2}$ s$^{-1}$, \citealt{hill2011}). The spectral energy
 distribution of these two components is plotted as a blue and
 magenta thick line in Fig.~\ref{fig:spectrum}, respectively, together
 with the IR/optical/UV (red thick line; \citealt{demartino2013}) and
 the radio spectrum (black thick line; \citealt{hill2011}).

The cut-off in the $\gamma$-ray energy band is well reproduced by
$\gamma_{max}=10^4$. On the other hand, the shape of the X-ray power
law results from the sum of the synchrotron and SSC contribution in
that energy band (with the former contributing for more than two
thirds of the emission), and is reasonably modelled by using an index
$\alpha=2.4$ for the electron energy distribution.  Keeping fixed
these two parameters, and choosing $\bar{B}=5\times 10^6$ G, we found a
good modelling of the observed X-ray and $\gamma$-ray spectra for an
electron density of $n_e=12\times10^{17}$ cm$^{-3}$.  The spectral
energy distribution so obtained is plotted in Fig.~\ref{fig:spectrum}
as a black solid line, where the synchrotron and SSC components are
also drawn as a blue and magenta dotted line, respectively. The low
energy cut-off of the synchrotron spectrum is set at the energy
predicted by Eq.~\ref{eq:lowenbreak}.

In Fig.~\ref{fig:spectrum}, we also plot the inverse Compton spectrum
yielded by seed photons coming from the inner disk (assuming a inner
temperature of 100 eV and a truncation radius of 50 km) and the
bremsstrahlung spectrum (evaluated for a fully ionised plasma with an
electron density equal to the value determined above) as a red and
green dotted line, respectively. Their contribution to the total
spectrum can be then safely neglected, for the parameters considered
in this case, according to expectations.

Similarly, the thermal X-ray output of the inner parts of the
truncated accretion disk is expected to be at most of the same order
of the synchrotron emission, at energies of few tenths of keV. To show
this, we used the model developed by \citet{gierlinski1999} to model
the spectrum of an disk in the accreting high state, truncated at 50
km, with an inner disk temperature of 100 eV (see
Sec.~\ref{sec:iccem}), a high inclination of 75$^{\circ}$ (as possibly
indicated by the dips and flares observed from the source), a
hardening factor of 2, and interstellar absorption by a column with
density $N_{\rm H}=10^{21}$ cm$^{-2}$ (see cyan dashed line in
Fig.~\ref{fig:spectrum}). This component would have a 0.5--10 keV flux
not larger than a factor of few with respect to the upper limit which
can be set on the presence of a disk thermal component during the
January 2009 observation ($2\times10^{-13}$ erg cm$^{-2}$ s$^{-1}$).
Considering the large uncertainties on the thermal emission of an disc
truncated by a propellering magnetosphere, and the scattering of most
of the disk soft X-ray photons by the electron cloud (see red dashed
line in Fig.~\ref{fig:spectrum}), we then consider that the non
detection of a thermal component at soft X-rays is still compatible
with our description.

The IR/optical/UV spectrum was modelled by \citealt{demartino2013} as
the sum of the contributions of the companion star ($T_2=4600\pm250$
K) and of the outer part of the accretion disc ($T_h=12800\pm600$ K;
$R\approx 10^5$ km). Considering also the low energy cut-off of the
synchrotron spectrum at few eV, the emission coming from the
transition layer is not expected to contribute directly to more than
ten per cent to the IR/optical/UV output. However, UV flares are
simultaneous to X-ray flares, despite they have a lower amplitude, by
a factor of two, indicating that the emission in the two bands are
closely related \citep{demartino2010,demartino2013}. This is still
compatible with our model if the UV emission is assumed to be due to
reprocessing in the outer rings of the disc of the X-ray emission
generated at the inner disc boundary, something already suggested by
\citealt{demartino2013}.

The emission model that we have developed for the radiation coming
from the transition layer cannot explain the optically thin/flat radio
spectrum ($F_{E}\sim\nu^{p}$, with $p=-0.5\pm0.6$) observed by ATCA at
frequencies of 5.5 and 9 GHz \citep{hill2011}. It is evident from
Eq.~\ref{eq:lowenbreak} that a similar emission has to come from a
region of a larger size, with a correspondingly lower electron density
and magnetic field strength. For instance, in order to obtain a break
of the synchrotron spectrum below 5.5 GHz, and assuming a dipolar
decay of the field as $B(r)\propto r^{-3}$, the emitting region should
be 100 times larger and have a density lower by four orders of
magnitude. These properties would be anyway compatible with an
emitting region pertaining to the binary system.

\subsection{Implications for \src }
\label{sec:results}

The model parameters that we obtained can be used together with the
relation derived in Sec.~\ref{sec:propeller} and \ref{sec:sed}, to get
model-dependent constraints on the system parameters. The cut-off
energy of the electron distribution ($\gamma=10^4$) and the magnetic
field strength at the interface ($\bar{B}=5\times10^6$ G), are related
by Eq.~\ref{eq:gammamax} to the acceleration parameter $\xi$, and the
ratio between the synchrotron and the SSC luminosity, $f$. For the
considered parameters, we have $f=1.98$ and an acceleration parameter
$\xi\simeq0.07$ (see Table~\ref{table}).

Eq.~\ref{eq:solution} relates the strength of the field at the
interface to the radiated luminosity, the spin period, the fastness
and the elasticity parameter. Setting $P_{2.5}=1$, and the luminosity
of $1.49\times10^{35}$ erg s$^{-1}$ as obtained from our modelling,
gives a fastness parameter of $1.45$ for the elastic propeller case
(larger than the critical value of $1.21$ needed for mass ejection,
\citealt{perna2006}), a magnetic dipole moment of $3.1\times10^{26}$ G
cm$^3$ (Eq.~\ref{eq:dipole}) and a mass ejection rate of
$6.3\times10^{15}$ g s$^{-1}$ (Eq.~\ref{eq:ejection}), compatible with
the typical mass accretion rates observed from NS in LMXB. For these
parameters, the inner disc radius would be placed at $R_{\rm
  in}\simeq40$ km, with an acceleration region of transverse section
$R_{\rm t}\simeq0.1$ km.
 
We also varied the strength of the magnetic field at the interface
$\bar{B}$ to study the range of parameters which lead to a good fit of
the observed spectral energy distribution, and compatible with the the
propeller model developed in Sec.\ref{sec:propeller}. Decreasing
$\bar{B}$ while keeping fixed the spin period, increases the range of
values of the elasticity parameters which can provide a solution above
the value of the critical fastness. For instance, a value of
$\bar{B}_6=4$ gives a propeller solution for $\beta>0.4$, while
$\bar{B}_6=3$ is allowed for $0.1>\beta>0$. A similar effect is
obtained by increasing the value of the spin period. A too low value
of the field (e.g. $\bar{B}_6=1$) at the interface is not formally
compatible with the propeller model we developed; in this case, in
fact, a solution of Eq.~\ref{eq:solution} is found for $\omega<1.03$,
which is below the critical fastness for any elasticity parameter. For
such values the field would then more likely produce an accretion
state than a propeller. On the other hand, when $\bar{B}$ is increased
above $6\times10^6$ G, the volume of the acceleration zone needed to
keep the contribution of the SSC photons comparable to that yielded by
synchrotron photons, decreases uncomfortably, while the acceleration
parameter goes above a value of 0.1. At the same time the magnetic
dipole moment increases above $10^{27}$ G cm$^3$, which is also
unlikely for a NS in a LMXB. We therefore conclude that an interface
magnetic field strength in a relatively narrow range of
$3$--$6\times10^6$ G best reproduces the observed spectra and is
compatible with the theoretical expectations for a NS in a LMXB,
ejecting mass in a propeller state. This range of field imply an
  acceleration parameter in the range 0.04-0.08, to give a comparable
  emission in the two components (see Eq.~\ref{eq:gammamax}). The
parameter values of a sample set of models are given in
Table~\ref{table}.

\subsection{Hadrons acceleration}
\label{sec:hadrons}

So far, we have not analysed possible acceleration of hadrons,
followed by subsequent interactions with material in the disc, pion
decay, and gamma-ray production. This is an alternative that may in
principle require consideration.
In our scenario, where particles get accelerated in the transition
zone, use of Eq.~\ref{eq:tauacc} and \ref{eq:tautravel}, which give
the timescale of acceleration and escape, applied to the case of
protons, imply that the maximum proton acceleration would happen for a
Lorentz factor of 10$^4$, i.e., an energy of 10 TeV.  We do not see
the outcome of these accelerated protons, which would produce photons
up to a few hundreds GeV, since the gamma-ray spectrum is severely cut
at a few GeV.  (For electrons, the maximum energy of the population is
limited not by escape losses, but by synchrotron and SSC, which can be
instead neglected in the case of protons, as they are a factor
$(m_p/m_e)^2\approx4\times10^6$ less efficient, see
Eq.~\ref{eq:synchro}) One could in principle entertain that protons of
the highest energies will penetrate the inner structure of the
accretion disc, and interacting there would produce a photon of few
100 GeV which would very likely interact as well, being absorbed
\citep[see, e.g.,][]{bednarek1993}.  We cannot discard this scenario a
priori, without detailed computations, as a possible contributor to
the total SED yield.
What may seem less likely is the picture in which the protons are
accelerated in an electrostatic gap out of the transition zone, which
then impact the accretion disk in a sort of beam-meets-target
phenomenon \citep{cheng1989}. This scenario has been explored for some
high-mass X-ray binaries, like A0535+26 (earlier claimed as a possible
EGRET source) by \citet{romero2001}, and \citet{anchordoqui2003};
albeit the model has not been confirmed by Fermi-LAT or at higher
energies \citep{acciari2011}. In our case, assuming first we are in
accretion phase, not in propeller, and using the formulae in
\citealt{romero2001}, the acceleration potential would produce one
order of magnitude less voltage than for A0535+26, and the current
flowing in the disc would also be one order of magnitude less.

\section{Discussion}
\label{sec:disc}

Being the only LMXB with a proposed bright persistent $\gamma$-ray
counterpart discovered so far, {\src} is an extremely intriguing
source. Though, the nature of the compact object hosted by this system
is still uncertain. In this paper, we proposed that the system hosts a
neutron star in a propeller state, developing a model to reproduce the
observed X-ray and $\gamma$-ray properties of the source.  Before
discussing the implication of the model we presented in this paper, we
briefly summarise other possibilities suggested to explain at least
partly the rich phenomenology observed from {\src}.

\subsection{Can it be a rotational-powered pulsar?}
\label{sec:discpulsar}
Coherent pulsations were not detected from {\src} in the X-ray and
radio band, nor from its proposed $\gamma$-ray counterpart. The upper
limit on the X-ray pulse amplitude set by \citep[][between 15 and 25
  per cent in the 0.5--10 keV band]{demartino2013} are larger than
those observed from many accreting millisecond pulsars \citep[see,
  e.g.][and references therein]{patruno2012c}, and should not be
considered particularly constraining as to whether an accreting pulsar is
present in the system.

On the other hand, the non detection of radio pulsations reported by
\citet{hill2011} on the basis of Parkes 1.4 GHz observations indicates
that if {\src} harbours a rotation powered pulsar, either its pulsed
emission is not beamed towards the Earth, or it is scattered and
absorbed by matter engulfing the system. 
Indeed, the similarity of the $\gamma$-ray spectrum of the proposed
counterpart of {\src} to those observed by Fermi from several
$\gamma$-ray pulsars, led \citet{hill2011} to suggest that the system
may host a similar radio quiet/faint $\gamma$-ray pulsar. However, we
note that the X-ray emission of {\src} ($L_X=6.5(1)\times10^{33}$ erg
s$^{-1}$ in the 2--10 keV band) is larger by orders of magnitude
than that expected and usually observed from rotation powered pulsars
with a low mass companion. In fact, similar systems usually harbour a
weakly magnetised ($B\simeq10^8$--$10^9$ G) pulsar, spun up to a
millisecond spin period by a previous phase of mass accretion
\citep[see, e.g.][]{bhattacharya1991}. Expressing the spin-down power
of a pulsar as
\begin{equation}
\dot{E}\approx\frac{\mu^2}{c^3}\left(\frac{2\pi}{P}\right)^4,
\end{equation}
with $\mu$ magnetic dipole moment, and $P$ the spin period of the
neutron star, and considering a $\eta=L_X/\dot{E}\leq 10^{-3}$
efficiency of the conversion of the spin down power in X-ray
luminosity \citep[e.g.][]{becker2009}, typical parameters observed in
millisecond pulsars ($P\approx\mbox{few}$ ms; $\mu\approx10^{26}$ G
cm$^{-3}$) yield:
\begin{equation}
\dot{L_X^{psr}}\simeq 1.5\times10^{31}\,\eta_{-3}\,\mu_{26}^2\,P_{2.5}^{-1}\,\mbox{erg s}^{-1}.
\end{equation}
Here, $\eta_{-3}$ is the X-ray conversion efficiency in units of
$10^{-3}$. Indeed, the brightest rotation powered millisecond pulsars
in X-rays have luminosities of $\simeq10^{33}$ erg s$^{-1}$
\citep{cusumano2003,webb2004,bog2011}.  Similarly, pulsars which
showed a transition between rotation and accretion powered states, PSR
J1023+0038 \citep{archibald2009}, and IGR J18245--2452
\citep{papitto2013}, were observed at an X-ray luminosity of
$\approx10^{32}$ erg s$^{-1}$ during their rotation powered activity
\citep{archibald2010,bog2011b}.
To match a similar value, {\src} should be then closer than 0.6 kpc,
spinning rather rapidly and/or being particularly young. We note that
a distance $1.4$--$3.6$ kpc is suggested by the spectral shape of the
colder thermal component detected in the optical band by
\citet{demartino2013}. 

Also, the observation of optical emission lines would not seem to
easily fit a scenario with a rotation powered pulsar, as in a similar
state the pulsar wind would be expected to sweep the entire Roche Lobe
of the neutron star from the matter transferred by the companion star
\citep{ruderman1989}. For similar reasons we consider the rotation
powered pulsar scenario is at least improbable, as {\src} should be by
far the brightest rotation powered pulsar, without pulsations being
detected, and with a disk surviving the intense radiation pressure
which would be implied by such a high spin down power.

\subsection{Can it be an accreting black hole?}

The non-detection of coherent pulsations from {\src} and its short
term X-ray variability leaves open the possibility that an accreting
black hole in a low hard state is present in the system
\citep{saitou2009,demartino2013}. In such case, the radio emission
would be originated in a compact jet. Indeed, the largely uncertainty
of the radio spectrum (a $F_{\nu}\propto\nu^{-\alpha}$ power-law
spectrum, with $\alpha=0.5\pm0.6$) makes it compatible with the
typical flat spectrum observed from compact jets ($\alpha\approx
0$). As noted by \citet{demartino2013}, the observed ratio between the
radio luminosity at 9 GHz and the X-ray luminosity in the 3--9 keV
X-ray band place it slightly under-luminous in the radio band with
respect to the correlation observed for black hole binaries
\citep{gallo2003}, while it seems  over-luminous with respect
to accreting neutron stars in the hard state
\citep{migliari2006}. However, the relatively bright gamma-ray
emission seems difficult to reconcile with a black-hole scenario. So
far, GeV emission has been detected only from two systems hosting a
black hole; a weak emission recently detected from Cyg X-1
\citep{malyshev2013} --albeit this has not been confirmed in
subsequent analysis by the {\it Fermi}-LAT collaboration--, and a
brighter but transient emission observed from Cyg X-3
\citep{tavani2009,fermi2009}. In both cases, however, according to
leptonic models the high energy emission is related to up-scattering
of the dense photon field emitted by the massive companion star and/or
the disc, which are hardly important contributors in the case of
{\src}. Indeed, no gamma-ray emission has been reported so far from
the many black holes with a low-mass companion star known. Thus we
also consider this scenario unlikely.

\subsection{A neutron star in propeller}

In this paper we have proposed a propellering neutron star scenario
for {\src}. Our model is based on the assumption that in a similar state a
population of electrons can be accelerated to relativistic energies at
the interface between the disc and the magnetosphere, following the
suggestion  put forward by \citet{bednarek2009,bednarek2009b}. 
He applied a similar model to the case of slowly rotating ($P\ga 10$
s) accreting NS in HMXB ($B_{\rm NS}\sim10^{12}$ G, $\mu\sim 10^{30}$
G cm$^{3}$; \citealt{bednarek2009b}), as well as to NSs with
super-critical magnetic field at their surface ($B_{\rm NS} \sim
10^{14}$ G, $\mu\sim 10^{32}$ G cm$^{3}$), harboured in binary system
with a massive companion star \citep{bednarek2009}.  This concept was
later developed by \citet{torres2012,papitto2012} to explain the
multi-wavelength phenomenology of LS I 61 303; and has found support in
the recent discovery of the super orbital variability of the gamma-ray
emission from the system \citep{ackermann2013}.

While the
surface magnetic field of a typical NS in a LMXB is lower by more
than four orders of magnitude than the much more intense fields of NS
in HMXB or in magnetars, the radius at which the matter in-flow is
truncated in a NS-LMXB system is much lower. The field at the
magnetospheric interface of a NS in a LMXB, like that hypothesised for
{\src}, is then up to three orders of magnitude larger in this case
(Eq.~\ref{eq:dipole}), and as a consequence also the power available
to accelerate electrons (Eq.~\ref{eq:engain}).

For typical parameters of a system like the one considered here, the
cooling of this electron population takes place mainly through
synchrotron interaction with the magnetic field permeating the
interface, and through inverse Compton losses due to the interaction
between the electrons and the synchrotron photons. The dominance of
self-synchrotron Compton emission is not usually encountered in
systems. We found that LMXB in a propeller state could be prone to
this situation.
Inverse Compton
losses given by the interaction with the radiation field emitted by
the disc, and by the companion star, represent in fact a contribution
which is orders of magnitude lower. The same holds for
bremsstrahlung losses.  

As both the dominant cooling channels are strongly dependent on the
strength of the magnetic field, this quantity has a crucial influence
on the value of the maximum energy yielded to electrons. We showed
that for typical parameters of a propellering NS in LMXB
($\mu\approx\mbox{few}\times10^{26}$ G cm$^{3}$, $P\approx\mbox{few
  ms}$, $R_{\rm in}\ga 50$ km), and an acceleration parameter in
  the range 0.01--0.1, a maximum energy of few GeV is naturally
obtained, compatible with the high energy cutoff observed from the
$\gamma$-ray counterpart of {\src}. At the same time, if the emission
region is compatible with the size of the magnetosphere-disc interface
($R_{\rm in}\ga 50$ km; $R_{t}\approx\mbox{km} $), the synchrotron
self Compton emission will give rise to an emission with an overall
output comparable to that yielded in X-rays by the synchrotron
emission. A similar model is therefore able to explain
semi-quantitatively the peculiar spectral energy distribution of
{\src} at high energies. On the other hand, synchrotron absorption in
the emission region predicts a low energy cut-off at few eV.
According to our model, the emission arising at the magnetospheric
interface cannot take into account the radio emission observed from
the source, which should therefore come from a larger region with a
lower electron density.

The model we presented may also explain the observational features
recently observed from the transitional pulsar PSR
J1023+0038. Following the disappearance of radio pulsations, the onset
of an accretion state has been recently reported for this otherwise
rotational-powered system \citep{stappers2013}. Evidences supporting
the formation of an accretion disc has been obtained from optical
observations \citep{halpern2013}, similar to those which let
\citet{archibald2009} to conclude that the source was in an accretion
state between 2000 and 2001 \citep[see
  also][]{wang2009}. Simultaneously, the source brightened by more
than an order of magnitude in X-rays to an average level of
$L_X(0.5$--$10\,\mbox{keV})\simeq2.5\times10^{33}$ erg s$^{-1}$ (with
variations by a factor 10 on timescales of few tens of seconds
\citealt{patruno2013b,kong2013,papitto2013b}), and by at least a
factor of five in gamma-rays ($L_{\gamma}(>100\,\mbox{MeV})\ga
5\times10^{33}$ erg s$^{-1}$, \citealt{stappers2013}), with respect to
the rotation-powered phase characterised by an active radio-pulsar
\citep{archibald2010,tam2010,bog2011b}. The observed value of the
X-ray luminosity strongly suggests that the system has entered a
propeller state, possibly alternating with a rotational-powered state
on short timescales of tens of seconds \citep{patruno2013b}. The model
we presented can be taken as a plausible interpretation of the
comparable emission observed from this system in X-rays and
gamma-rays, alternative to a scenario in which the gamma-ray emission
is due exclusively to residual periods of activity as a
rotational-powered pulsar.

 In the model presented here we considered a neutron star in a purely
 ejecting propeller state, even if \citet{bednarek2009,bednarek2009b}
 argued that the electron acceleration at the magnetospheric interface
 can take place both if the neutron star is effectively accreting the
 in-flowing mass down to its surface, and if it is instead
 propellering mass away. Such a choice was made as in such conditions
 the interface between the field and the disc is expected to be highly
 turbulent and magnetised, thus favouring the acceleration of
 electrons through a Fermi process. Even if the accretion of a
 fraction of the in-flowing matter (like in the trapped state studied
 by \citealt{dangelo2010,dangelo2012} and applied by
 \citealt{patruno2009b,patruno2013} to interpret properties of a few
 accreting pulsars) is in principle possible, the low X-ray flux
 observed from the source and the absence of a detected thermal
 component in the soft X-ray band set a limit on the accretion rate.
 Even assuming that the total X-ray output of the source is powered by
 accretion, this would not take place at a rate larger than $\simeq
 2\times10^{-12}$ M$_{\odot}$ yr$^{-1}$, i.e. $10^{-4}$ times the
 Eddington rate.

According to our model, the observed X-ray and gamma-ray emission (as
well as the matter outflow) are ultimately powered by the in-fall of
matter down to the disc truncation radius, and by the energy deposited
by the rotating magnetosphere (see Eq.~\ref{eq:mdot}). The observed
X-ray flares/dip pairs could be then produced by a sudden increase of
the rate of mass in-fall, caused by inhomogeneities of the disc
accretion flow, and a subsequent re-fill of the starved parts of the
disc, similar to the interpretation given by
\citealt{demartino2010,demartino2013}. On the other hand, the observed
UV emission is larger by an order of magnitude than the contribution
of the synchrotron component of our model at those energies. To
explain the simultaneity among flares and dips observed at X-rays and
UV energies, we should then conclude that the latter emission is
mainly due to reprocessing in the outer parts of the disc, of the
emission at higher energies produced close to the magnetospheric
boundary.

As a concluding remark, we note that positional correspondences
between persistent and transient X-ray binaries and $\gamma$-ray
sources detected by Fermi and Agile seem intrinsically rare
\citep[e.g.][and references
  therein]{ubertini2009,sguera2011,li2012b}. \citet{maselli2011}
searched for positional correspondence between sources of the second
Palermo BAT catalogue \citep{cusumano2010} and the first Fermi
catalogue \citep{abdo2010} and found only 15 galactic sources, among
which only two are LMXB, {\src} and SLX 1735--269. Intriguingly, also
the latter is a faint persistent X-ray source, and on this basis was
classified by \citet{intzand2007} as a candidate ultra-compact X-ray
binary. The increase in sensitivity obtained by Fermi as its mission
progresses will help shedding light on the possibility that more
$\gamma$-ray LMXB candidates exist; at least, in those cases where the
accretion state is such that a steady gamma-ray emission exist.

\section*{Acknowledgments}

Work done in the framework of the grants AYA2012-39303, as well as
SGR2009-811, and iLINK2011-0303. AP is is supported by a Juan de la
Cierva Research Fellowship. DFT was additionally supported by a
Friedrich Wilhelm Bessel Award of the Alexander von Humboldt
Foundation. AP thanks D. De Martino for illuminating discussions.

 \bibliographystyle{mn2e}
 \bibliography{biblio}

\end{document}